\documentclass[twocolumn,aps,pre,groupedaddress]{revtex4-2}
\usepackage{mystyle} 
\usepackage{terms_in_models} 

\date{\today} 

\newcommand{\PAPER}{paper} 
\newcommand{\STUDY}{study} 

\newcommand{\red}[1]{{\textcolor{red}{ #1}}} 

\begin{document}

\title{Phase transition  of parallelizability in assembly systems} 
\author{Ikumi Kobayashi}
\author{Shin-ichi Sasa}
\affiliation{\KyotoPhys} 

\begin{abstract} 
We propose a phase transition on the feasibility of efficient parallel assembly.
 By introducing the {\IndexName} that measures how efficiently the parallel assembly works, the parallelizable phase is defined by its positive value.
The {\ParaUnpara} transition is then identified by the non-analytic change in the parallel efficiency from a positive value to zero.  
We present two analyzable models to demonstrate this phase transition in the limit of infinite system size.
\end{abstract}

\maketitle

\section{Introduction}
In industry and applied science, there are often situations where a large number of parts are assembled to make a complex product.
Typical examples are automotive assembly and polymer synthesis.
Recently,  the advancement of nanotechnology \cite{poole2003introduction} has made it possible to assemble nanoscale objects into desired structures \cite{bishop2009nanoscale, nykypanchuk2008dna, koh2007strategies, grzybowski2017dynamic}. 
Furthermore,  control of polymer sequences \cite{badi2009sequence, szymanski2018exploring} and  assembly of colloidal particles \cite{chen_directed_2011, velev2006chip, juarez2012feedback, tang2016optimal, li2011colloidal, grzybowski2017dynamic} have been vigorously studied.

We use the term parallelization to describe simultaneously assembling subunits and then combining them to complete the final product.
Parallelization increases assembling efficiency.
The concept of parallelization has been studied in computer science \cite{greenlaw_limits_1995, arora2009computational}.
 Some problems can be efficiently solved by parallel computing, whereas others cannot \cite{Note1}. 
\fnt{1}{This is called the NC versus P problem in computational complexity theory.}
Inspired by these studies, we explore analogous concepts in physical assembly work.
Specifically, we aim to determine under what conditions efficient parallel assembly is feasible.

The feasibility of parallelization qualitatively changes the time required for assembly. 
When the number of parts $L$ becomes large,
the $L$-dependence of the number of {\DAN} $d$ required for assembly is crucial. 
For example, when assembling hundreds to thousands of parts,
the assembly time is drastically different depending on whether $d = \order{\log{L}}$ or $d = \order{L^\alpha}$.

The feasibility of parallel assembly can be clarified by introducing the {\IndexName} $\eta$. 
The {\IndexName} $\eta$ is defined as the ratio of the minimum number of steps $\log_2 L$ required to assemble $L$ parts to the actual number of steps $d$ taken for assembly.
For instance, imagine the assembly of a 2-mer is achieved by combining two monomers, followed by the combination of 2-mers to form 4-mers, and so on.
Under such a fully parallelized assembly, we have $d=\log_2 L$ and $\eta=1$. 
Conversely, in the case of sequential assembly where components are added one by one, we have $d=L-1$ and $\eta= \log_2 L / (L-1)$, which goes to zero in the limit of $L\to\infty$.

In this {\PAPER}, we propose a phase transition called \textit{\TransitionName},
where {\ParaUnpara} phases are characterized by {\IndexName}.
That is, when a system parameter is continuously changed, {\IndexName} exhibits a transition from a positive value to zero in the limit of infinite system size.
We demonstrate this phase transition by presenting two analyzable models.
In the first model,  the {\MODELone}, {\oneD} chains are assembled in the smallest number of {\DAN}. 
In the second model,  the {\MODELtwo},  a final product is assembled through random bonding reactions.
We introduce the {\IndexName} to measure how efficiently the parallel assembly works.
Then, we exactly show that both models exhibit the {\TransitionName}.

\section{Setup of  the {\MODELone}}
Let us consider the assembly work of connecting $\Lone$ different parts to create a single chain.
An external operator tries to perform the most efficient parallel assembly possible.
However, the components do not always fit together.
Which pairs of states can be combined is predetermined and does not change during the assembly process.
 The  {\MODELone} idealizes such a situation.

We consider the assembly of {\oneD} chains of length $L$.
To precisely specify the geometric structure of the states, we use graph theory notations and terminologies.
We denote as $G_{i,j}$ a path graph in which vertices from $i$ to $j$ are connected in order:
\ARRAY{l}{1}{
G_{i,j} = (V,E)\\
V = \SET{\FromTo{i}{j}}\\
E = \SETc{(v,v+1)}{v = \FromTo{i}{j-1}}.
}
See \Figref{Gij} for the illustration of $G_{i,j}$.

\EPS[width=5.0cm]{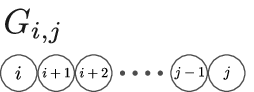}{We denote as $G_{i,j}$ a path graph in which vertices from $i$ to $j$ are connected in order. For simplicity, we assume that there are edges between touching vertices and omit describing the edges.}

The {\GsName} $G$ is a path graph of length $L${\ie} \EQ{G = G_{1,L}.}
The set of possible states $S$ is the entirety of the connected subgraphs of $G${\ie} \EQ{S = \SETc{G_{i,j}}{1\leq i \leq j \leq L}.}
The {\MsName} $M$ is the set of states with a single vertex{\ie} \EQ{M = \SETc{G_{i,i}}{1\leq i \leq L}.}
The case $\Lone=7$ is shown in \Figref{model1_states_7}.

\EPS[width=8.6cm]{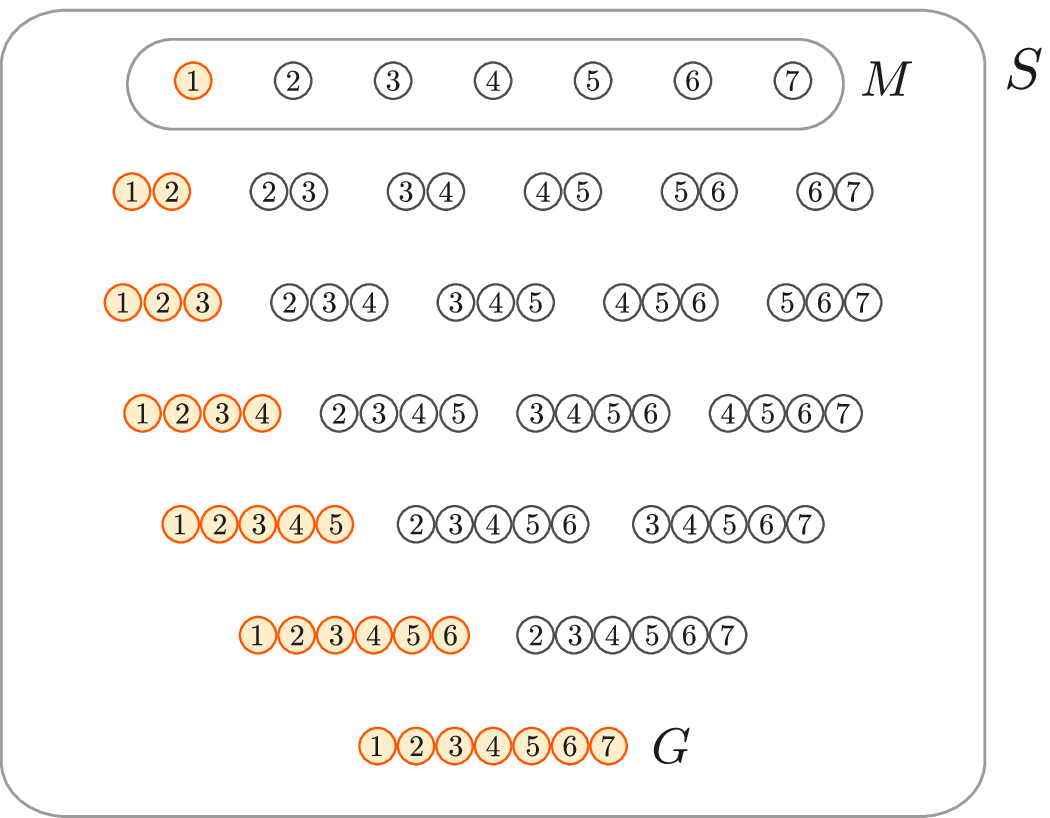}{All elements of $S$ in the case $\Lone=7$. The {\GsName} $G$ is shown at the bottom. The set of possible states $S$ is the entirety of the connected subgraphs of $G$. The {\MsName} $M$ is shown at the top.}

Each state $s \in S$ is either active (filled circles in \Figref{model1_states_7}) or inactive (open circles in \Figref{model1_states_7}).
This active/inactive distinction represents the bonding properties of the state with other states.
That is, active states can always combine with other states, 
while inactive--inactive pairs can combine with probability $p$.
Note that randomness is introduced as a quenched disorder.
We determine the set of allowed bondings $\hat{R}$ probabilistically according to the following rules:\\

For each tuple $i,j,k\,\,(1\leq i \leq j < k \leq \Lone)$,
\nITEM{5pt}{
\item if either $G_{i,j}$ or $G_{j+1,k}$ is active, $(G_{i,j}, G_{j+1, k}) \in \hat{R}$ with probability 1;
\item if both $G_{i,j}$ and $G_{j+1,k}$ are inactive, $(G_{i,j}, G_{j+1, k}) \in \hat{R}$ with probability $\pone$.
}
These rules are illustrated in \Figref{reaction}.
Let us assume that the activity is carried over to the post-bonding state.
That is, the product in case 1 is active and the product in case 2 is inactive.
In addition, we suppose that only $G_{1,1}$ is active in $M$.
From the initial condition that only $G_{1,1}$ is active in $M$ and the propagation rule of the active state, we have
\EQ{G_{i,j} = \twoCASES{\textrm{active}}{(i=1)}{\textrm{inactive}}{(i\neq 1)}.}

\EPS[width = 8.0cm]{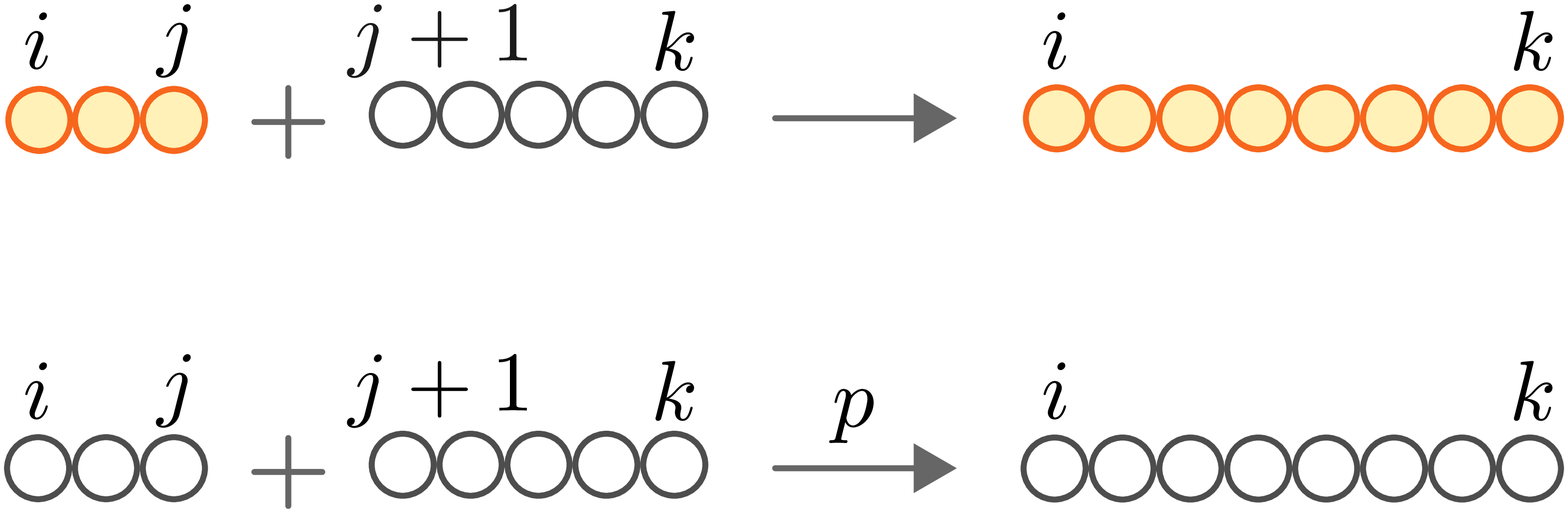}{
\SCHEMA{the decision procedure for the set of possible bondings $\hat{R}$}
For each pair of states $(G_{i,j}, G_{j+1, k})$, 
(i) if either $G_{i,j}$ or $G_{j+1,k}$ is active $(i=1)$, $(G_{i,j}, G_{j+1, k}) \in \hat{R}$ with probability 1;
(ii) if both $G_{i,j}$ and $G_{j+1,k}$ are inactive $(i>1)$, $(G_{i,j}, G_{j+1, k}) \in \hat{R}$ with probability $\pone$.
By making this determination for every pair $(G_{i,j}, G_{j+1, k})$ of graphs, we determine the set $\hat{R}$ probabilistically.
}

Let us introduce the {\IndexName} to measure how well the parallelization is working in this system.
We call diagrams like \Figref{model1_d_7} {\TREE}s.
In the {\TREE} shown in \Figref{model1_d_7}, a chain of length $L=7$ is assembled with four {\DAN}.
The number of {\DAN} means the maximum distance from the upmost states to the bottom state.
We denote by $d(T)$ the number of {\DAN} of the {\TREE} $T$.

\EPS[width=5.0cm]{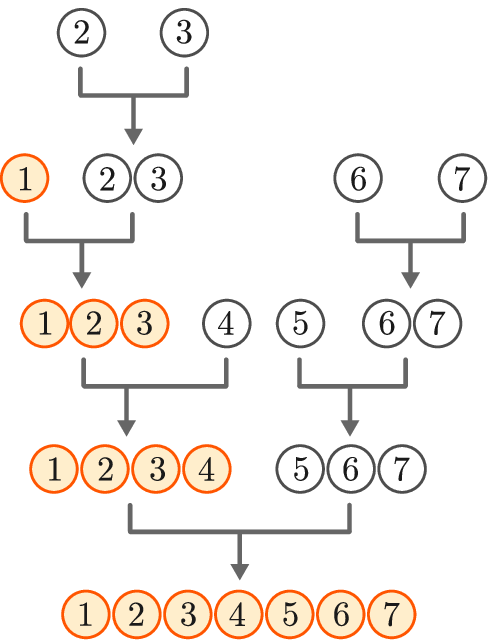}{
 Example of an {\TREE} generating $G_{1,7}$. In this case, $d(T)=4$.
{\cTREE}s are binary trees that express the building process of a state.
The product is placed at the bottom.
The elements of $M$ are placed at the top.
}

\NOTATION{$\done$}{the least number of {\DAN} required to assemble the {\GsName} $G$}{\ie}
\EQ{\done \DEF \MINc{T\in \hat{U}_G}{d(T)},}
where $\hat{U}_G$ is the set of all {\KA} {\TREE}s generating $G$.
Note that not all {\TREE}s are necessarily {\KA}.
For example, if $(G_{5,5}, G_{6,7}) \notin \hat{R}$, the {\TREE} in \Figref{model1_d_7} is {\HUKA}.
Because the set of allowed bondings $\hat{R}$ is probabilistically determined, $\done$ is also a random variable.

The {\IndexName} $\etaone$ is defined as
\EQ{\etaone \DEF \FRAC{\log_2{\Lone}}{\AVE{\done}},}
where $\Lone$ is the total number of vertices and $\dave$ is the average minimum number of {\DAN}.
We can use the quantity $\etaone$ to measure the efficiency of parallel assembly in this system because it satisfies the following two properties.
First, $\etaone$ satisfies the normalization condition $0\leq \etaone \leq 1$.
Second, we can determine the feasibility of efficient parallel assembly in the limit of infinite system size by checking whether $\lim_{L\toinf}\eta$ is positive or zero.
If the assembly process is sufficiently parallelized and $\dave \sim \log L$, then $\lim_{L\toinf}\eta$ is positive.
 In contrast, if the parallelization breaks down and $\dave$ grows faster than $\log L$, then $\lim_{L\toinf}\eta$ becomes zero.

\section{ Results for the  {\MODELone}}
We first display the numerical results in \Figref{model1_result}.
Although the data suggest the existence of a phase transition,
it is quite difficult to perform the numerical calculation for a larger system. 
Nevertheless, we have a rigorous proof for the existence of the
phase transition (see also Appendix) when $\Lone$ becomes infinite.
We can show 
\EQ{\lim_{\Lone \toinf} \etaone = 0  \label{model1_result_a}}
for $0\leq \pone < 1/4$, and 
\EQ{\lim_{\Lone \toinf} \etaone \neq 0  \label{model1_result_b}}
for $3/4 \leq \pone \leq 1$. Thus, the analyticity of $\etaone$
is broken at a point $\pcone$ satisfying $1/4 \leq \pcone < 3/4$.
This result is illustrated in \Figref{model1_result}.
The breaking of the analyticity of $\etaone$ allows us to
identify the {\ParaUnpara} phases without arbitrariness.
In other words, the region of $\pone$ satisfying $\lim_{\Lone \toinf}\etaone \neq 0$ is identified as a parallelizable phase and the region of $\pone$ satisfying $\lim_{\Lone \toinf}\etaone = 0$ as an unparallelizable phase.

\EPS[width=8.6cm]{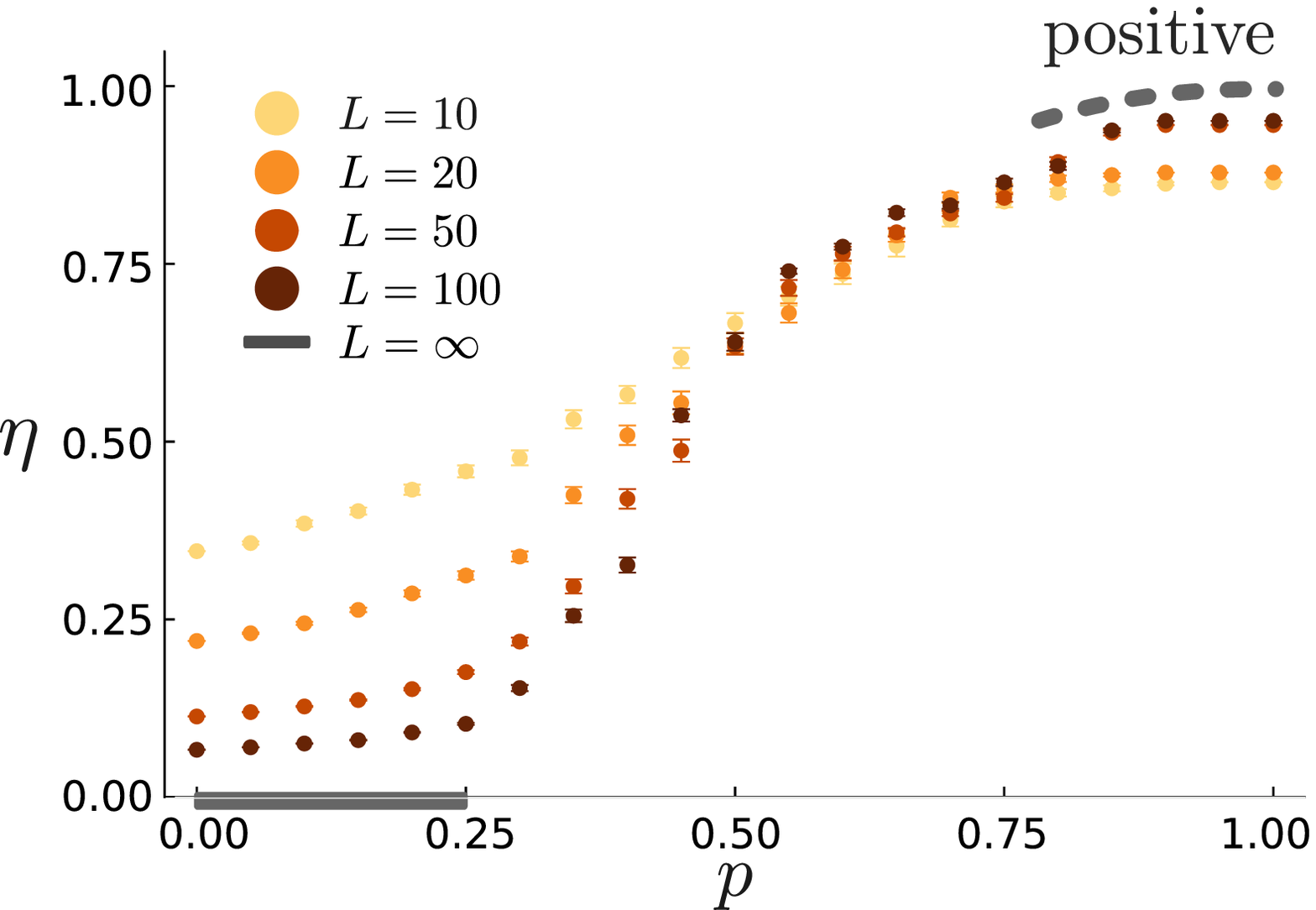}{
\SUCHI{$\Lone =$ \fourAND{10}{20}{50}{100} in  the  {\MODELone}}
For each $\pone$, we took 100 samples and calculated the mean and standard deviation of $\etaone$.
 The schematic of {\IndexName} $\etaone$ in the limit of infinite system size is overlaid.
The dashed line represents qualitative behavior.
A rigorous analysis shows that the analyticity of $\etaone$ is broken at a point  $\pcone$ satisfying $1/4 \leq \pcone < 3/4$.
}

Here, we briefly outline the proof of \twoEqsref{model1_result_a}{model1_result_b}.
A rigorous proof is given in Appendix.
The essence of the proof of \Eqref{model1_result_a} which represents the unparallelizable phase is that when $p$ is small, it is difficult to construct large, inactive states. 
The number of {\TREE}s  generating a state with $n$ vertices is given by the Catalan number, which asymptotically increases as $4^n$. 
The realization probability of each {\TREE} is $p^{n-1}$. 
From the balance of these two factors, it can be understood that when $p$ is smaller than 1/4, there is a high possibility that there are no ways to construct large inactive states. 
Therefore, when $p$ is smaller than 1/4, it is necessary to connect inactive states one by one using active states, and it is found that the assembly cannot be completed in $\order{\log L}$ steps. 
The obtained upper limit of the parameter of the unparallelizable phase, 1/4, originated from the asymptotic behavior of the Catalan number.

The essence of the proof of \Eqref{model1_result_b} which represents the parallelizable phase is that when $p$ is large, there is a high possibility of the existence of relatively unbiased {\TREE}s. 
For example, considering an {\TREE} with no bias such that both children of a state with $n$ vertices have $n/2$ vertices, we have $d = \log_2 L$. 
In contrast, considering a very biased {\TREE} such that the children of a state with $n$ vertices have $n-1$ and 1 vertices, we have $d = L-1$. 
In the proof of \Eqref{model1_result_b}, it is shown by mathematical induction on the number of vertices that when $p$ is greater than 3/4, at least one relatively unbiased {\TREE} is likely to exist. 
Therefore, it is found that when $p$ is greater than 3/4, there is a high possibility that there is at least one {\TREE} with $d = \order{\log L}$. 
In contrast to the case of the parallelizable phase, the obtained lower bound, 3/4, does not have a clear origin. 
It is possible to slightly improve the lower bound by increasing the number of the base cases of the induction.

\section{Setup of  the  {\MODELtwo}}
Instead of optimizing the {\TREE}s, we study typical pathways of stochastic evolution of assembly in the second model.
This model is interpreted as a mean-field version of the first model.
To simplify the analysis, inactive states in  the  {\MODELone}
are further classified into neutral states and passive states in the  {\MODELtwo}.
That is, each component takes one of three states: active, neutral, or passive.
In the {\MODELone}, once it is determined that two states $s_1$ and $s_2$ cannot combine, they will never combine during the assembly process.
The {\MODELtwo} incorporates this effect by assuming that passive states never combine with each other.

The stochastic assembly rule is as follows. 
Initially, there are one active and $(\Ltwo-1)$ neutral components.
The assembly of the {\BUNSHI}s proceeds in a repetition of the following two steps: 
(i) randomly pair two {\BUNSHI}s as possible  \cite{Note4}; 
(ii) for each pair, perform the following bonding reaction:
an active component can bond with any other component,
 a neutral component can bond with inactive components with probability $\ptwo$,
 and a passive component cannot bond with passive components.
 If a neutral component fails to bond, the component becomes passive.
We define the set of operations (i) and (ii) as a single round.
\fnt{4}{When the total number of {\BUNSHI}s is odd, leave the extra one
and do nothing until the next round. }

 The {\MODELtwo} is expressed symbolically by denoting 
active, neutral, and passive  components as  $\SEED$, $\NEW$,
and $\OLD$, respectively.
Operation (ii) is then written as the following  set of  chemical reactions: 
\EQ{\SEED + \NEW \to \SEED, \label{model2_bonding1}}
\EQ{\SEED + \OLD \to \SEED , \label{model2_bonding2}}
\EQ{\NEW + \NEW \to \twoCASES{\NEW}{\WP \ptwo}{\OLD + \OLD}{\WP 1-\ptwo}, \label{model2_bonding3}}
\EQ{\NEW + \OLD \to \twoCASES{\NEW}{\WP \ptwo}{\OLD + \OLD}{\WP 1-\ptwo} , \label{model2_bonding4}}
\EQ{\OLD + \OLD \to \OLD + \OLD. \label{model2_bonding5}}
This procedure is illustrated in \Figref{model2_round}.
The active state is necessary to ensure that the assembly process can always be executed. 
Even if all states become passive, the assembly process can be completed by the reaction represented by \Eqref{model2_bonding2}.

\EPS[width=7.5cm]{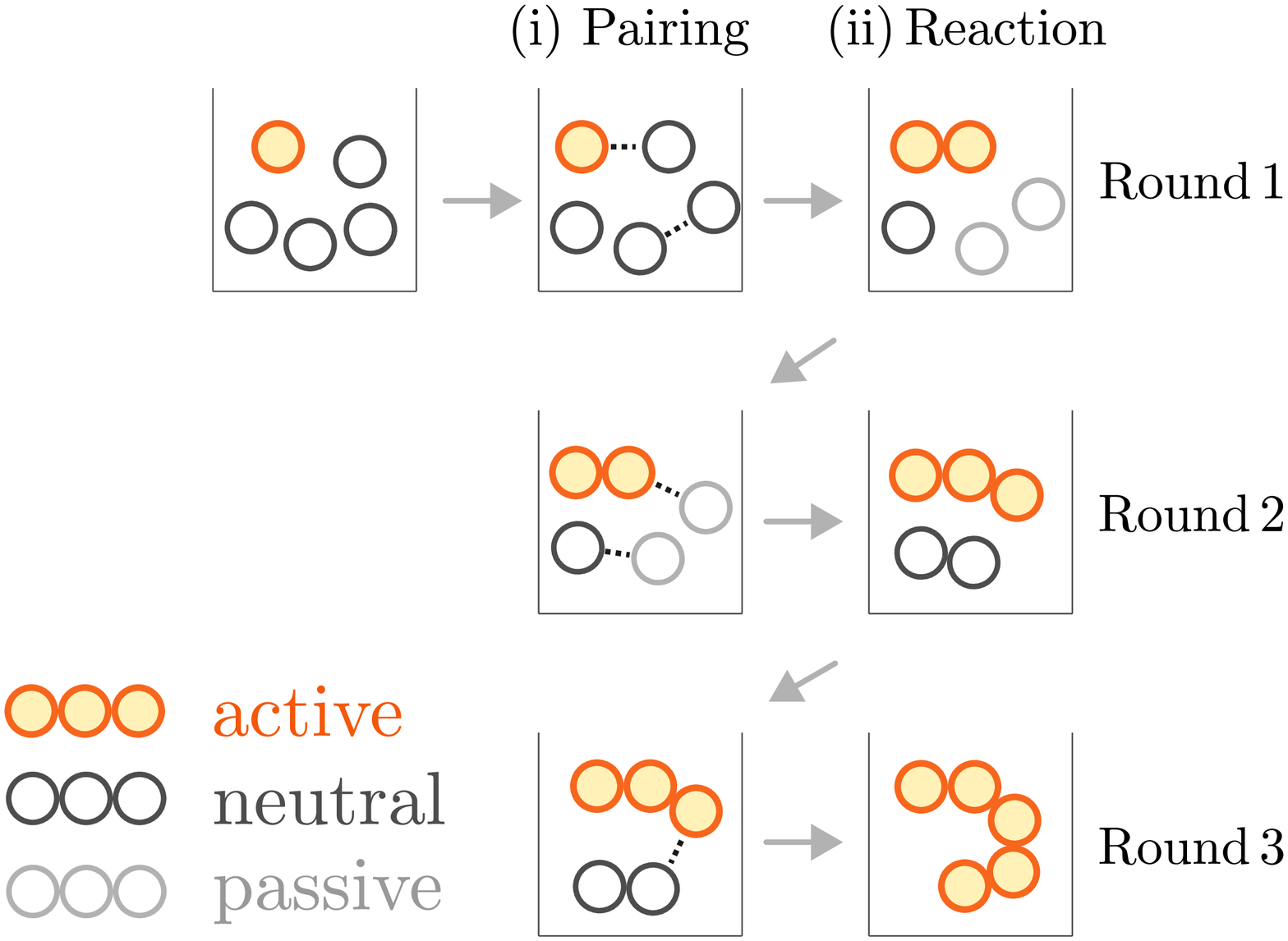}{
\SCHEMA{the assembly of five {\BUHIN}s in three rounds ($\Ltwo = 5, \dtwo = 3$)}
Initially, there are one active and $(\Ltwo-1)$ neutral components.
A single round consists of the following two steps: 
(i) Randomly pair two {\BUNSHI}s as possible,
(ii) For each pair, perform the reactions shown in \seqEqref{model2_bonding1}{model2_bonding5}.
 The quantity $\dtwo$ is the number of rounds until there is only one {\BUNSHI}. 
}

We measure the number of rounds $\dtwo$ until all {\BUHIN}s are connected.
The assembly of five {\BUHIN}s in three rounds is shown in \Figref{model2_round}.
The bonding reactions occur probabilistically, and the number of rounds required to connect all the {\BUHIN}s varies from trial to trial.
Therefore, $\dtwo$ is a random variable.

We introduce the  {\IndexName} $\etatwo$ to characterize the feasibility of the efficient parallel assembly.
The {\IndexName} $\etatwo$ is defined as
\EQ{\etatwo \coloneqq \FRAC{\log_2{\Ltwo}}{\AVE{\dtwo}},}
where $\Ltwo$ is the number of {\BUHIN}s and $\AVE{\dtwo}$ is the  average number of required rounds.
 In a way similar to the first model, $\etatwo$ satisfies the following two properties.
First, $\etatwo$ satisfies  the  normalization condition $0\leq \etatwo \leq 1$.
Second, we can determine the feasibility of efficient parallel assembly by checking whether $\etatwo$ is positive or zero in the limit $L \toinf$.

\section{ Results for the  {\MODELtwo}}
The simulation results are shown in \Figref{model2_result}.
These graphs suggest that a discontinuous transition exists at a point $\pctwo$. 
Indeed, for this model, we prove that the {\IndexName} $\etatwo$
 has  a discontinuous transition when $\Ltwo$ becomes infinite.
Quantitatively, we show
\EQ{\lim_{\Ltwo \toinf} \etatwo = 0  \label{model2_result_a}}
{for $0\leq \ptwo < \pctwovalue$, and 
\EQ{\lim_{\Ltwo \toinf} \etatwo \neq 0  \label{model2_result_b}}
for $\pctwovalue < \ptwo \leq 1$.
This implies that the parallel efficiency $\etatwo$ is discontinuous
at $\pctwo = \pctwovalue$.

\EPS[width=8.6cm]{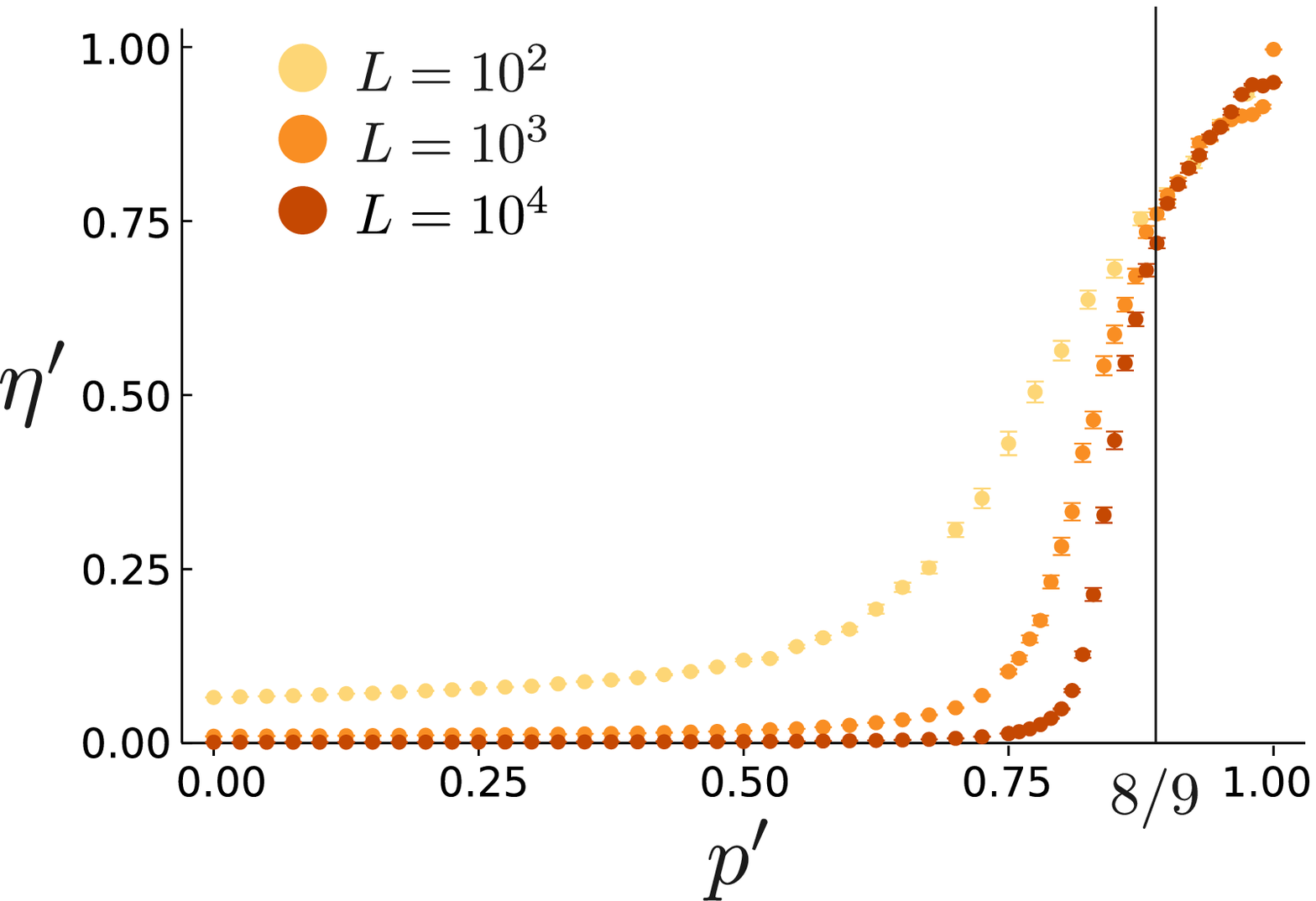}{
\SUCHI{the {\MODELtwo}}
For each $\ptwo$, we took 100 samples and calculated the mean and standard deviation of $\etatwo$.
The data points approach the line $\ptwo = \pctwovalue$ as $\Ltwo$ increases.}

The proof is the following.
Let \threeAND{$\seed{n}$}{$\new{n}$}{$\old{n}$} be the population of \threeAND{$\SEED$}{$\NEW$}{$\OLD$}
in the $n$th round, respectively. 
The total population of
{\BUNSHI}s is \EQ{\tot{n} = \seed{n} + \new{n} + \old{n} \label{tot_def}.}
Because the population of $\SEED$ does not change through the reactions,
$\seed{n} = 1$ always holds.
Assuming that $\Ltwo$ is large enough, we analyze the behavior of the expected values of \threeAND{$\new{n}$}{$\old{n}$}{$\tot{n}$} ignoring terms of $\order{1}$.
\NOTATION{$\AVE{\hat{S_n}}$}{the expected number of $\OLD - \OLD$ pairs in the $n$th round}.
We obtain
\EQ{\AVE{\hat{S_n}} \simeq \FRAC{\tot{n}}{2} \qty(\FRAC{\old{n}}{\tot{n}})^2 \label{S_eq},}
where the symbol $\simeq$ represents an approximation ignoring the terms of $\order{1}$ \cite{Note5}.
\fnt{5}{
Consider a random variable $\hat{\sigma_i}$ such that $\hat{\sigma_i} = 1$ when the $i$th-pair is $\OLD - \OLD$ and $\hat{\sigma_i} = 0$ otherwise.
Then $\AVE{\hat{S_n}} \simeq \AVE{\sum_{i=1}^{\tot{n}/2} \hat{\sigma_i}} = \tot{n}/2 \cdot \AVE{\hat{\sigma_i}}$, and we obtain \Eqref{S_eq}.
}
For each pair  described  by \twoOR{\Eqref{model2_bonding3}}{\Eqref{model2_bonding4}}, $\NEW$ is generated $\WP  \ptwo$.
Therefore, we obtain
\ARRAY{lll}{2}{
\new{n+1}
&=& \qty(\FRAC{\tot{n}}{2} - \AVE{\hat{S_n}}) \ptwo\\
&\simeq& \FRAC{\tot{n}}{2}\qty(1 - \qty(\FRAC{\old{n}}{\tot{n}})^2) \ptwo.
}
 Because  the number of {\BUNSHI}s is halved in the reaction that produces $\SEED$ or $\NEW$, we obtain
\EQ{\tot{n+1} = \tot{n} - (\seed{n+1}+\new{n+1}) \simeq \tot{n} - \new{n+1}.
}
By setting \EQ{q_n \DEF 1 - \qty(\FRAC{\old{n}}{\tot{n}})^2 \label{q_def},}
we obtain
\EQ{\new{n+1} \simeq \FRAC{\tot{n}}{2}q_n \ptwo, \label{new_n+1}}
\EQ{\tot{n+1} \simeq \qty(1-\FRAC{q_n \ptwo}{2}) \tot{n}. \label{tot_n+1}}
\SUBS{\threeEqsref{tot_def}{new_n+1}{tot_n+1}}{\Eqref{q_def}}
\EQ{q_{n+1} \simeq \FRAC{q_n \ptwo \qty(1 - \frac{3}{4}q_n \ptwo)}{\qty(1- \frac{1}{2}q_n \ptwo)^2}. \label{q_eq}}

The discrete dynamical system given by \Eqref{q_eq} exhibits a {\snbunki} as the parameter $\ptwo$ changes.
This bifurcation structure is shown in \Figref{SNbunki}.
When $\ptwo$ is less than $\pctwovalue$, only $q_n = 0$ is a stable fixed point.
The assembly cannot be completed with $\dtwo = \order{\log \Ltwo}$ because $\tot{n+1}/\tot{n} \simeq 1$. 
As a result, $\etatwo$ becomes zero in this case.
In contrast, when $\ptwo$ is greater than $\pctwovalue$, a new stable fixed point appears in $0 < q_n < 1$.
The total number of {\BUNSHI}s decreases exponentially because $0 < \tot{n+1}/\tot{n}<1$.
As a result, $\dtwo = \order{\log \Ltwo}$ and $\etatwo$ takes a positive value in this case.

\EPS[width=8.6cm]{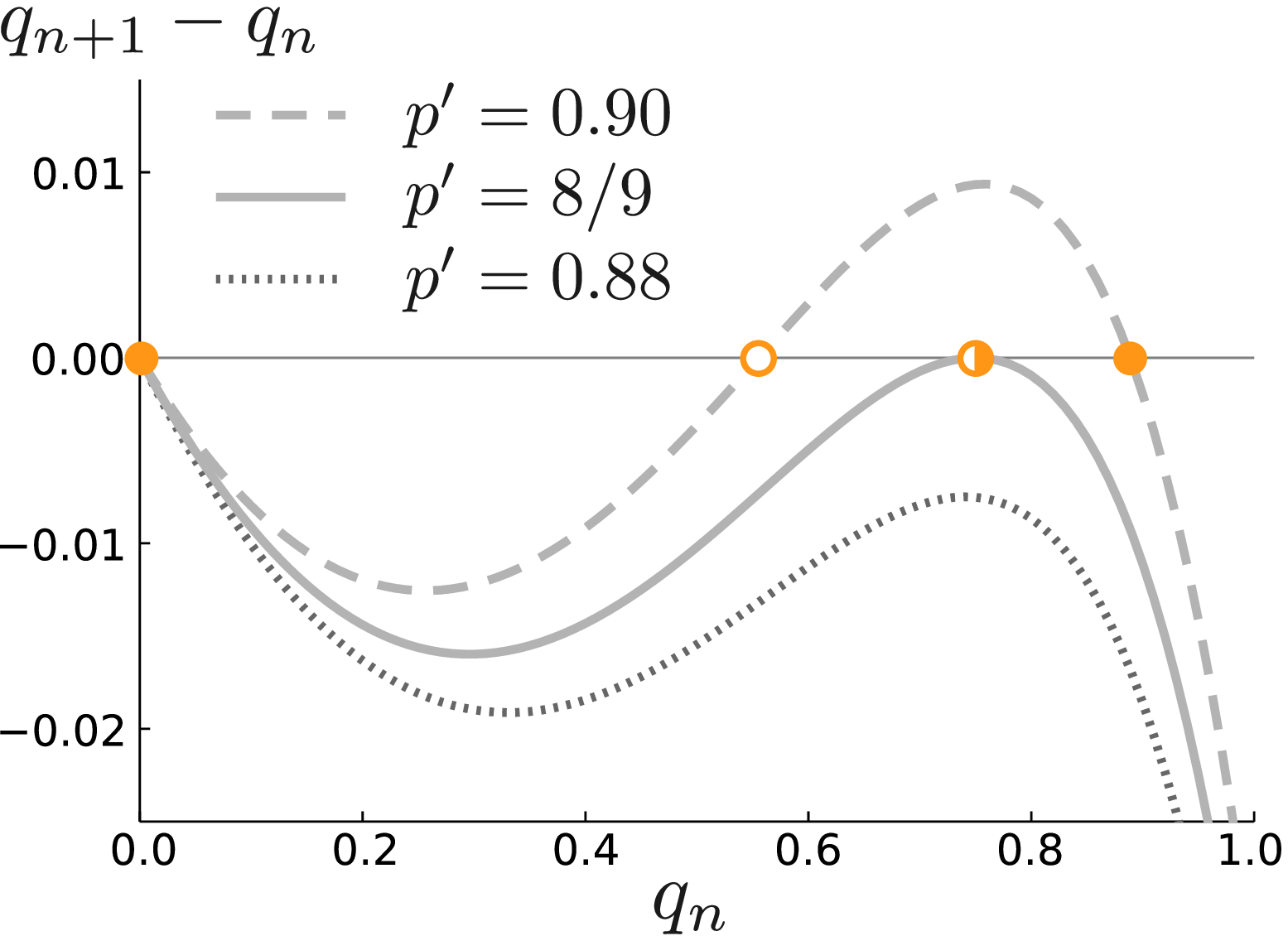}{
{\SNbunki} exhibited by the discrete dynamical system of \Eqref{q_eq}.
When $\ptwo$ is less than $\pctwovalue$, only $q_n = 0$ is a stable fixed point.
When $\ptwo$ is greater than $\pctwovalue$, a new stable fixed point and an unstable fixed point appear in $0 < q_n < 1$.
}

\section{Discussion}
The two models presented in this {\PAPER} characterize in-principle and realistic parallelizability, respectively.
In the  {\MODELone}, the pathway with the least number of {\DAN} is chosen after considering all possible {\TREE}s.
Therefore, $\etaone$ characterizes whether efficient parallel assembly is feasible in principle.
The equivalent situation would be bottom-up manufacturing of industrial products, where the manufacturing process is well-designed and optimized in advance.
In the  {\MODELtwo}, $\Ltwo$ {\BUHIN}s are randomly paired and combined, and the number of rounds until all {\BUHIN}s are connected is measured.
Therefore, $\etatwo$ characterizes whether efficient parallel assembly is realistically possible.
The equivalent situation would be chemical synthesis, where the molecules randomly collide.

The {\IndexName} defined in this {\STUDY} is related to the complexity of molecular structures.
The minimum number of {\DAN} $\done$ is essentially the same as the molecular assembly index (MA) defined in the literature \cite{marshall2021identifying, marshall2017probabilistic, marshall2019quantifying}.
 It may be possible to extend this {\STUDY} to classify the complexity of molecules using {\IndexName}.

The model analyzed in this {\STUDY} can be considered a variation of several known models.
By excluding the single active unit, the {\MODELone} can be considered as a form of directed percolation in (1+1) dimensions \cite{hinrichsen2000non}. 
Therefore, findings in directed percolation may be used to estimate the transition point in this model.
The {\MODELtwo} can be related to the cluster merging process described by the Smoluchowski equation \cite{ziff1982critical}. 
In this process, the number of clusters exhibits exponential decay, which corresponds to the parallelizable phase in this {\PAPER}.
The {\MODELtwo} can also be viewed as one special case of stochastic chemical reactions or reaction-diffusion systems \cite{van1992stochastic, gardiner1985handbook}.
One such model is the activated random walk, which consists of active particles A and sleeping particles S \cite{Levine_2021}.
Discussing parallelizability in general chemical reaction systems is an important future task.

We present possible future directions.
As a practical direction, this {\STUDY} can apply to actual industrial production processes.
{\cParaUnpara} transitions would emerge in connection to the success rate $p$ of each process during the assembly of complex structures.
Applying the method of this {\STUDY} may make it possible to calculate the threshold success rate of the elementary process to achieve efficient parallel assembly.

As a theoretical direction, this {\STUDY}  could lead to methods of classifying chemical reaction systems using {\IndexName}. 
 Chemical reaction systems are classified according to the number of steady states or the number of conserved quantities \cite{feinberg2019foundations}.
 Extending this {\STUDY} may make it possible to add another axis ({\ParaUnpara}) to the classification of chemical reaction systems.

The model discussed here may be realized in chemical reaction systems.
In organic synthetic chemistry, chemical reactions such as living radical polymerization \cite{moad2008toward, braunecker2007controlled} and multicomponent reactions \cite{kakuchi2014multicomponent, kakuchi2019dawn} are studied.
Such reaction systems could correspond directly to the model analyzed in this {\STUDY}.

\section{Conclusion}
In this {\PAPER}, we proposed a phase transition on the feasibility of efficient parallel assembly.
We demonstrated the {\ParaUnpara} transition through two models.
We can consider some extensions of the models.
For example, the {\MODELone} assumes that all reaction probabilities between inactive states are $p$, but this could be extended to depend on the internal composition to resemble a real chemical reaction. 
It is an important future task to extend the model to make the theory more easily comparable to real experiments.

\THANKS{\fourAND{Ryohei Kakuchi}{Masato Itami}{Tomohiro Tanogami}{Yusuke Yanagisawa}}
\KAKENHIs{\threeAND{JP19H05795}{JP20K20425}{JP22H01144}}

\appendix \label{model1_proof} 
\section{Proof of the results}
In this section, we prove \twoEqsref{model1_result_a}{model1_result_b}.
Before we begin the proof, we define the {\TREE} more formally.

An {\TREE} $T$ generating a state $s\in S$ is a binary tree that satisfies the following four conditions:
\nITEM{5pt}{
\item Each vertex of $T$ is an element of $S$.
\item The root of $T$ is the state $s$.
\item For each vertex $G_{i,j}$ of $T$, the children are \twoAND{$G_{i,k}$}{$G_{k+1,j}$} ($i\leq k \leq j-1$).
\item Every leaf of $T$ is an element of $M$.}
The number of {\DAN} $d(T)$ is the height of the tree $T$ \cite{Note1}.
\fnt{1}{The height of a tree is the maximum distance from the root to the leaf.}
An example of an {\TREE} $T$ generating a state $s = G_{3,7}$ is shown in \Figref{tree}.
Note that {\TREE}s are defined not only for $G$ but also for every state $s\in S$.
We call an {\TREE} $T$ generating an inactive state \textit{{\SUBTREE}}.

\EPS[width=6.0cm]{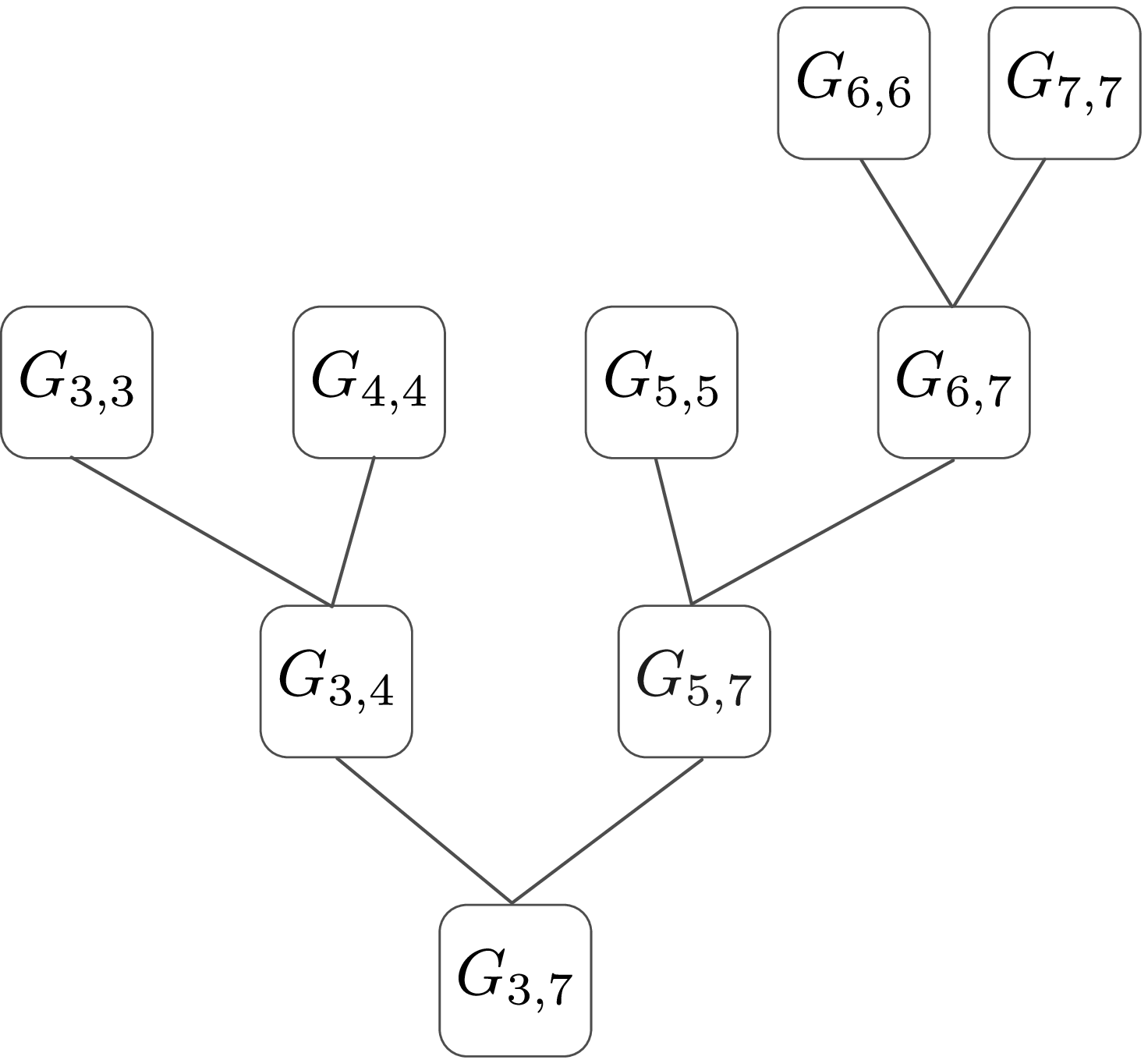}{
 Example of an {\TREE} $T$ generating a state $s = G_{3,7}$. Each vertex of $T$ is an element of $S$, that is, $G_{i,j}$.
The root of $T$ is the state $s$.
For each vertex $G_{i,j}$ of $T$, the children are \twoAND{$G_{i,k}$}{$G_{k+1,j}$} ($i\leq k \leq j-1$).
Every leaf of $T$ is an element of $M$, that is, $G_{i,i}$.
In this case, $d(T) = 3$.}

\NOTATION{two children of a state $s$}{\twoAND{$C_s^1$}{$C_s^2$}.
Such vertices \twoAND{$C_s^1$}{$C_s^2$} having the same parent are called siblings.}
We then introduce the following terms: 
\ITEM{5pt}{
\item An {\TREE} $T$ is \textbf{{\KA}} if all siblings $(C_s^1, C_s^2)$ in $T$ are included in $\hat{R}$
\item An {\TREE} $T$ is \textbf{{\HUKA}} if there exist siblings $(C_s^1, C_s^2)$ in $T$ that are not included in $\hat{R}$.}

\subsection{Unparallelizable phase \label{sequential_phase}} 
We show the proof of \Eqref{model1_result_a}.
When $p=0$, $\lim_{\Lone \toinf} \etaone = 0$ is trivial because $\dave = \Lone-1$ holds.
Thus, we consider the case $0 < \pone < 1/4$.

\subsubsection{Number of {\TREE}s \label{about_Catalan}}
\NOTATION{$a_n$}{the total number of {\TREE}s of a path graph with $n$ vertices}.
Focusing on the last step, we obtain the following {\ZENKASIKI}:
\ARRAY{lll}{2}{
a_1 &=& 1\\ 
a_n &=& \SUM_{i=1}^{n-1} a_{i} a_{n-i} \quad(n\geq 2).} 
This is the same as the {\ZENKASIKI} that defines the Catalan number.
Using the general terms of Catalan numbers \cite{koshy2008catalan}, we obtain
\EQ{a_n = \FRAC{(2(n-1))!}{n!(n-1)!}.}

This number has the following upper bound:
\EQ{a_n \leq 4^{n-1} \quad (n \geq 1). \label{Catalan_ub}}
\PROOFbyMI{\Eqref{Catalan_ub}}{
In the case $n=1$, $a_n \leq 4^{n-1}$ is true because $a_1 = 4^0 = 1$.}{
Assuming that $a_n \leq 4^{n-1}$ holds $(n = \FromOne)$, we obtain
\ARRAY{lll}{3}{
a_{n+1}
&=& \FRAC{(2n)!}{n!(n+1)!}\\
&=& \FRAC{2n(2n-1)}{n(n+1)}\times a_n\\
&=&\qty(4-\FRAC{6}{n+1})\times a_n\\
&<& 4\times 4^{n-1} = 4^n.
}}

\subsubsection{Number of {\SUBTREE}s \label{bubun}}
\NOTATION{$\calT_m$}{the set of all {\SUBTREE}s generating inactive states of size $m$}.
The number of the size $m$ inactive states is $(\Lone - m)$.
For each of them, there are
\EQ{a_m = \FRAC{(2(m-1))!}{m!(m-1)!}} {\SUBTREE}s (see \Appref{about_Catalan}).
Thus, we obtain
\EQ{\abs{\calT_m} = (\Lone-m)a_m.}

\subsubsection{Evaluation of $\done$}
\NOTATION{$A_T$}{a stochastic event that an {\TREE} $T$ is {\KA}}. 
Then, the following proposition holds.

\PROP{\label{d_eval}
\EQ{\AND_{m = n}^{\Lone-1} \AND_{T\in \calT_m} \overline{A_T} \qNARABAq \done \geq \FRAC{\Lone-1}{n-1} \quad (2\leq n \leq \Lone-1)}
}

\PROOF{
As shown in \Figref{clusters}, any {\TREE} $T$ generating $G$ is decomposed into the sequential bonding of {\SUBTREE}s to active states. 
\NOTATION{$K$}{the number of {\SUBTREE}s in $T$}.
\NOTATION{$m_k$}{the size of the state generated by the $k$th {\SUBTREE}} \seeFigref{clusters}.
 Because the premise of \Propref{d_eval} means that there is no way to generate an inactive state with more than $n$ vertices, $m_k \leq n-1$ holds.
Therefore, we obtain
\EQ{\Lone-1 = \SUM_{k=1}^{K} m_k \leq (n-1)K.}
Because the distance from the root $G$ to $G_{1,1}$ is $K$, $d(T) \geq K$ also holds. 
Therefore, we obtain
\EQ{d(T) \geq \FRAC{\Lone-1}{n-1}. \label{dT_lb}}
Because \Eqref{dT_lb} holds for any {\TREE} $T$, we obtain
\EQ{\done = \MINc{T\in \hat{U}_G}{d(T)} \geq \FRAC{\Lone-1}{n-1}.}
}

\EPS[width=5.0cm]{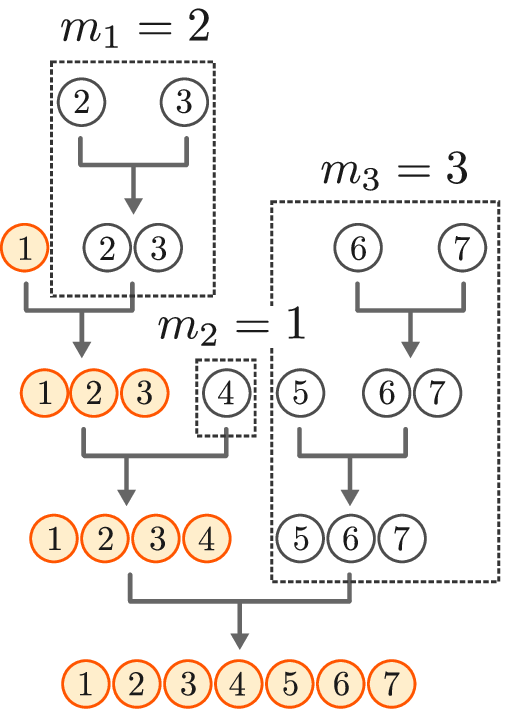}{
Any {\TREE} $T$ generating $G$ is decomposed into the sequential bonding of {\SUBTREE}s (dashed rectangles) to active states (filled circles).
This diagram shows the case $K=3$ and $d(T) = 4$.
}

 Assembling an inactive state with $m$ vertices requires $(m-1)$ bondings, which are independently realized with probability $\pone$.
Therefore, for any {\TREE} $T \in \calT_m$,
\EQ{\PROB{A_T} = p^{m-1},}
where $\PROB{A}$ represents the probability that stochastic event $A$ occurs.

\subsubsection{Evaluation of the probability}
Using the above preparation, we evaluate the probability as
\ARRAY{lll}{4}{\label{Prob_lb}
\PROB{\done \geq	\FRAC{\Lone-1}{n-1}} 
&\geq& \PROB{\AND_{m = n}^{\Lone-1} \AND_{T\in \calT_m} \overline{A_T}}\\
&=& \PROB{\overline{\OR_{m = n}^{\Lone-1} \OR_{T\in \calT_m} {A_T}}}\\
&=& 1- \PROB{\OR_{m = n}^{\Lone-1} \OR_{T\in \calT_m} {A_T}}\\
&\geq& 1- \SUM_{m=n}^{\Lone-1} \SUM_{T\in \calT_m} \PROB{A_T}\\
&=& 1- \SUM_{m=n}^{\Lone-1} \abs{\calT_m} p^{m-1}.
}
Note that \EQ{\PROB{\mrQ}\geq \PROB{\mrP}} holds when \EQ{\mrP \NARABA \mrQ} holds \seeFigref{inclusion}.
We used \twoAND{this relation}{\Propref{d_eval}} in the first line.
Furthermore, we used de Morgan's rule in the second line and Boole's inequality in the fourth line.
Boole's inequality, also known as the union bound, is an inequality given by
\EQ{\PROB{\OR_{i=1}^n E_i}\, \leq \,\, \SUM_{i=1}^n \PROB{E_i} ,}
where $E_i \,\,(i = \FromOneTo{n})$ represent arbitrary events which may not be independent.

\EPS[width=4.0cm]{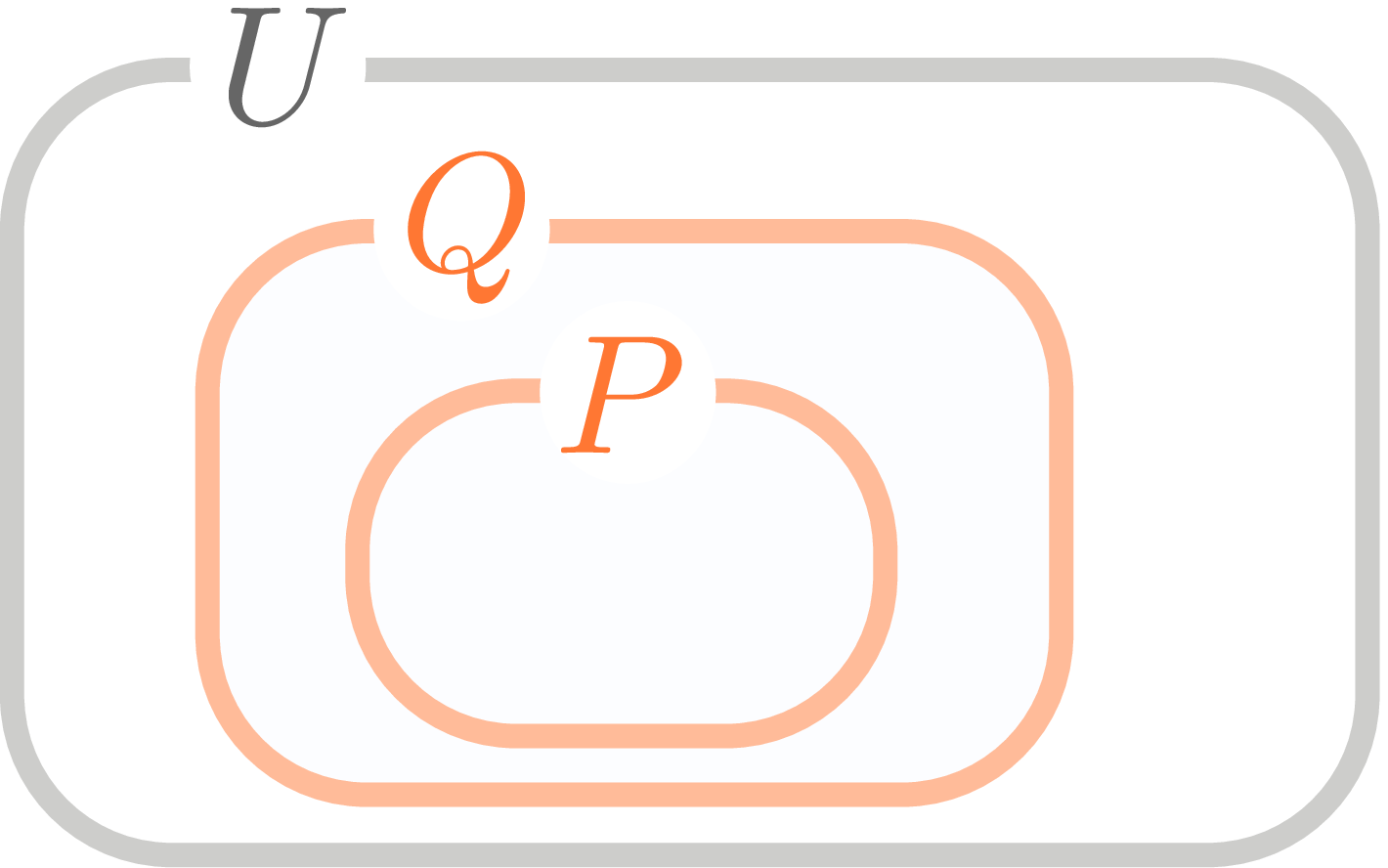}{
Diagram of the relationship between propositions and sets.
$\PROB{\mrQ}\geq \PROB{\mrP}$ holds when $\mrP \NARABA \mrQ$ holds.}

We further evaluate the sum as
\ARRAY{lll}{3}{\label{sum_ub}
\SUM_{m=n}^{\Lone-1} \abs{\calT_m} p^{m-1}
&=& \SUM_{m=n}^{\Lone-1} (\Lone-m) a_m p^{m-1}\\
&\leq& \SUM_{m=n}^{\Lone-1} (\Lone-m)(4\pone)^{m-1}\\
&<& \SUM_{m=n}^{\infty} \Lone(4\pone)^{m-1}\\
&=& \FRAC{\Lone(4\pone)^{n-1}}{1-4\pone},
}
where we used \twoAND{\Eqref{Catalan_ub} \seeAppref{about_Catalan} in the second line}{$\pone < 1/4$ in the fourth line}.

\subsubsection{Evaluation of $\dave$ and $\etaone$}
Let us define the integer
\EQ{n_0 \DEF 2 + \CEIL{\log_{4\pone}\qty(\FRAC{1-4\pone}{\Lone})} \label{M0}.}
Then, we obtain
\ARRAY{lll}{3}{\label{d_ub1}
\dave
&\geq& \PROB{\done \geq \FRAC{\Lone-1}{n_0-1}}\times \FRAC{\Lone-1}{n_0-1}\\
&>& \qty(1- \FRAC{\Lone(4\pone)^{n_0 -1}}{1-4\pone})\times \FRAC{\Lone-1}{n_0-1}\\
&\geq& (1-4\pone)\times \FRAC{\Lone-1}{n_0-1},
}
where we used \threeAND{Markov's inequality in the first line}{\twoEqsref{Prob_lb}{sum_ub} in the second line}{\twoAND{\Eqref{M0}}{$\CEIL{x}\geq x$} in the third line}.
Markov's inequality is an inequality given by
\EQ{\PROB{\abs{\hat{X}}\geq a}\, \leq \,\, \FRAC{\lrAVE{\abs{\hat{X}}}}{a},}
where $\hat{X}$ is a stochastic variable and $a>0$.

Using \Eqref{d_ub1}, we evaluate $\etaone$ as
\ARRAY{lll}{3}{
\etaone &\DEF& \FRAC{\log_2 \Lone}{\dave}\\
&\leq& \FRAC{\log_2 \Lone \cdot \qty(n_0 - 1)}{(1-4\pone)(\Lone-1)}\\
&<& \FRAC{\log_2 \Lone \cdot \qty(2+ \log_{4\pone}\qty(\FRAC{1-4\pone}{\Lone}))}{(1-4\pone)(\Lone-1)}.}

We thus obtain
\EQ{\lim_{\Lone \toinf} \etaone = 0\quad}
for $0 \leq \pone < 1/4$.

\subsection{Parallelizable phase} 
We show the proof of \Eqref{model1_result_b}.
\subsubsection{A strategy that enables logarithmic height assembly \label{log_assembly}}
We introduce another term:\\ 
An {\TREE} $T$ is \LOGTREE{\alpha}
if, for any bonding process $s_i + s_j\to s_k$ in $T$ \cite{Note5},
\EQ{\MIN{\abs{s_i}, \abs{s_j}} \geq \CEIL{\alpha \abs{s_k} \label{log_tree}},}
where $\abs{s}$ represents the number of vertices of the state $s$.
\fnt{5}{More formally, for a vertex $s_k$ in $T$ and its children $s_i$ and $s_j$, we define $s_i + s_j \to s_k$ as bonding process in $T$.}

\PROP{ \label{prop_d_log_tree}
If $T$ is \logtree{\alpha},
\EQ{d(T) \leq \FRAC{\log{\Lone}}{-\log(1-\alpha)} \label{d_log_tree}}
holds.}
\PROOF{
From \twoAND{\Eqref{log_tree}}{$\abs{s_i} + \abs{s_j} = \abs{s_k}$}, we obtain
\EQ{\MAX{\abs{s_i}, \abs{s_j}} \leq \abs{s_k}-\CEIL{\alpha \abs{s_k}} \leq (1-\alpha)\abs{s_k}.}
Using this inequality repeatedly, we obtain 
\EQ{1 \leq (1-\alpha)^{d(T)}\abs{G} = (1-\alpha)^{d(T)}L ,}
which is equivalent to \Eqref{d_log_tree}.
}

\subsubsection{Probability that this strategy is not available}
\NOTATION{$B_s$}{the stochastic event that all \logtree{\alpha} {\TREE}s generating a state $s$ are {\HUKA}}.
\twoNOTATION{$L_m$}{the subgraph consisting of the leftmost $m$ vertices of state $s$}{$R_{\abs{s}-m}$}{the subgraph consisting of the rightmost $\abs{s}-m$ vertices of state $s$} \seeFigref{LR}.
Focusing on the final step, we obtain
\MULT{B_s
\qDOUCHIq
\textrm{For all } m \,\,(\margin{\abs{s}} \leq m \leq \abs{s}-\margin{\abs{s}}), \\
\hspace{20mm} (L_m, R_{\abs{s}-m}) \notin \hat{R} \qquad \lor \\
\hspace{20mm} \qty{(L_m, R_{\abs{s}-m}) \in \hat{R} \,\, \land \qty(B_{L_m} \lor B_{R_{\abs{s}-m}})}.} 

\EPS[width=6.0cm]{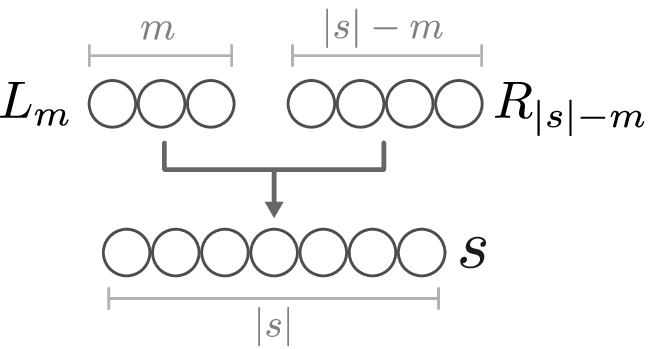}{
\twoNOTATION{$L_m$}{the subgraph consisting of the leftmost $m$ vertices of state $s$}{$R_{\abs{s}-m}$}{the subgraph consisting of the rightmost $\abs{s}-m$ vertices of state $s$}.}

Then, we obtain
\ARRAY{lll}{3}{ \label{Prob_B}
&&\PROB{B_s} \\
&=& \PROD_{m=\margin{\abs{s}}}^{\abs{s}-\margin{\abs{s}}} \qty((1-\tilde{\pone}) + \tilde{\pone} \, \PROB{B_{L_m} \lor B_{R_{\abs{s}-m}}})\\
&\leq& \PROD_{m=\margin{\abs{s}}}^{\abs{s}-\margin{\abs{s}}} \qty((1-\tilde{\pone}) + \tilde{\pone} \qty(\PROB{B_{L_m}} + \PROB{B_{R_{\abs{s}-m}}})),
}
where we defined $\tilde{\pone}$ as
\EQ{\tilde{\pone} = \twoCASES{\pone}{(L_m \textrm{ is inactive})}{1}{(L_m \textrm{ is active})}.}
In the second line, we used Boole's inequality.

Let us define
\EQ{Q_n \DEF \MAXc{\substack s\in S_n}{\PROB{B_s}} \label{QM_def},}
where $S_n = \SETc{s \in S}{\abs{s}=n}$.
Then, we obtain the following recursive inequalities for $Q_n$:
\ARRAY{lll}{3}{\label{Q_M}
Q_1 &=& 0\\
Q_n &\leq& \PROD_{m=\margin{n}}^{n-\margin{n}} \qty((1-\tilde{\pone}) + \tilde{\pone} \qty(Q_m + Q_{n-m})) \quad (n \geq 2).
}

\PROOF{
\NOTATION{$s^*\in S_n$}{the state for which $\PROB{B_s}$ is maximum}.
Applying \Eqref{Prob_B} to this $s^*$, we obtain
\ARRAY{lll}{3}{
Q_n
&=& \PROB{B_{s^*}}\\ 
&\leq& \PROD_{m=\margin{n}}^{n-\margin{n}} \qty((1-\tilde{\pone}) + \tilde{\pone} \qty(\PROB{B_{L_m}} + \PROB{B_{R_{n-m}}}))\\
&\leq&\PROD_{m=\margin{n}}^{n-\margin{n}} \qty((1-\tilde{\pone}) + \tilde{\pone} \qty(Q_m + Q_{n-m})),}
where we used \Eqref{QM_def} in the third line.
}

\subsubsection{Evaluation of $Q_n$}
\PROP{If $\pone \geq 3/4$ and $\alpha=1/6$, $Q_n \leq 1/4$ for all $n(=\FromOne)$.\label{Q_M_eval}}

\PROOFbyMI{\Propref{Q_M_eval}}{
Because $\alpha = 1/6$, $\margin{n} = 1$ for $n=2,3$. Then, we obtain
\ARRAY{ccccc}{2}{
Q_2 &\leq& (1-\tilde{\pone}) + \tilde{\pone}\cdot 2Q_1 &=& 1-\tilde{\pone}\\
Q_3 &\leq& \qty((1-\tilde{\pone}) + \tilde{\pone}(Q_1 + Q_2))^2 &\leq& (1-\tilde{\pone}^2)^2 .
}
Using $\tilde{\pone} \geq \pone \geq 3/4$, we obtain \twoAND{$Q_2\leq 1/4$}{$Q_3\leq 49/256 < 1/4$}. $Q_1 = 0 < 1/4$ also holds trivially.}{
Assume that $Q_1, Q_2, ... , Q_{n-1}\leq 1/4$ holds ($n=4,5,6,... $).
From \Eqref{Q_M}, we obtain
\ARRAY{lll}{3}{ \label{M+1}
Q_n 
&\leq& \PROD_{m=\margin{n}}^{n-\margin{n}} \qty((1-\tilde{\pone}) + 2\tilde{\pone}\cdot \FRAC{1}{4}) \\
&=& \PROD_{m=\margin{n}}^{n-\margin{n}} \qty(1-\FRAC{\tilde{\pone}}{2}) \\
&\leq& \qty(1-\FRAC{\pone}{2})^{n-2\margin{n}+1} \\
&\leq& \qty(1-\FRAC{\pone}{2})^3 \\
&\leq& \FRAC{125}{512} < \FRAC{1}{4},
}
where we used \fourAND{the {\KinouKatei} in the first line}{$\tilde{\pone}\geq \pone$ in the third line}{$n-2\margin{n} +1 \geq 3 \,\,(n\geq 4)$ in the fourth line}{$\pone \geq 3/4$ in the fifth line}.}

Substituting $Q_n\leq 1/4$ into \Eqref{Q_M} and using $\tilde{\pone}\geq \pone, \CEIL{x} < x+1$, we obtain the evaluation of $Q_n$:
\EQ{Q_n \leq \qty(1-\FRAC{\pone}{2})^{n-2\margin{n}+1} < \qty(1-\FRAC{\pone}{2})^{(1-2\alpha)n-1} \label{QM_ub}.}

\subsubsection{Evaluation of $\dave$ and $\etaone$}
We evaluate $\dave$ by separately considering the following two cases:
\nITEM{5pt}{
\item There exists a {\KA} \logtree{1/6} {\TREE} of $G$; that is, $B_G$ is false.
\item There exists no such {\TREE} of $G$; that is, $B_G$ is true.
}

In the first case, we can use $d \leq \log{\Lone}/(-\log(1-\alpha))$ through \Propref{prop_d_log_tree}, where $\alpha$ is set to $1/6$ to simplify the appearance.
Even in the second case, we can use the inequality $d<L$, which always holds.

We then obtain
\ARRAY{lll}{3}{ \label{para_dave}
\dave
&<& \PROB{\NOT{B_G}} \FRAC{\log{\Lone}}{-\log(1-\alpha)} + \PROB{B_G} L\\
&\leq& \FRAC{\log{\Lone}}{-\log(1-\alpha)} + Q_\Lone \Lone\\
&<& \FRAC{\log{\Lone}}{-\log(1-\alpha)} + \Lone \qty(1-\FRAC{\pone}{2})^{(1-2\alpha)\Lone-1},
}
where we used \twoAND{$\PROB{\NOT{B_G}} \leq 1$ and \Eqref{QM_def} in the second line}{\Eqref{QM_ub} in the third line}.

We then evaluate $\etaone$ using \Eqref{para_dave} as
\ARRAY{lll}{3}{
\etaone &\DEF& \FRAC{\log_2 \Lone}{\dave}\\
&>& \FRAC{\log_2 \Lone}{\FRAC{\log{\Lone}}{-\log(1-\alpha)} + \Lone \qty(1-\FRAC{\pone}{2})^{(1-2\alpha)\Lone-1}},
}
where the right side goes to $-\log_2(1-\alpha)$ in the limit $\Lone \toinf$.
Substituting $\alpha=1/6$ into the result, we obtain
\EQ{\lim_{\Lone \toinf} \etaone \neq 0}
for $3/4 \leq \pone \leq 1$.

\bibliographystyle{myaps}
\bibliography{para2022}

\end{document}